\documentclass[conference]{IEEEtran}
\usepackage[T1]{fontenc}
\usepackage{amsmath,bm,amsfonts,amssymb,graphicx,color,multirow,setspace,xfrac,comment}
\usepackage{booktabs}
\usepackage{epstopdf}
\usepackage{mathtools}
\usepackage{lipsum}
\usepackage{cite}
\usepackage{citesort}
\usepackage{float}
\usepackage{balance}
\begin{document}
\title{Real-Time Deployment Aspects of C-Band and Millimeter-Wave 5G-NR Systems\vspace{-4pt}}
\author{\IEEEauthorblockN{Mansoor Shafi\IEEEauthorrefmark{1},	
                         Harsh Tataria\IEEEauthorrefmark{2},
  						 Andreas F. Molisch\IEEEauthorrefmark{3},
						 Fredrik Tufvesson\IEEEauthorrefmark{2}, and Geoff Tunnicliffe\IEEEauthorrefmark{1}}
  \IEEEauthorblockA{\IEEEauthorrefmark{1}Spark New Zealand, Wellington and Auckland, New Zealand}
  \IEEEauthorblockA{\IEEEauthorrefmark{2}Department of Electrical and Information Technology, Lund University, Lund, Sweden}
  \IEEEauthorblockA{\IEEEauthorrefmark{3}Department of Electrical Engineering, University of Southern California, Los Angeles, CA, USA}
  \IEEEauthorblockA{e--mail: \{mansoor.shafi, geoff.tunnicliffe\}@spark.co.nz, \{harsh.tataria, fredrik.tufvesson\}@eit.lth.se, and molisch@usc.edu\vspace{-9pt}}}
\maketitle

\begin{abstract}
Fifth-generation (5G) new radio (NR) deployments are being rolled out in both the C--band (3.3 - 5.0 GHz) and millimeter-wave (mmWave) band (24.5 - 29.5 GHz). For outdoor scenarios, the C--band is expected to provide wide area coverage and throughput uniformity, whereas the mmWave band is expected to provide ultra-high throughput to dedicated areas \emph{within} the C-band coverage. Due to the differences in the frequency bands, both systems are expected to be designed with different transmit and receive parameters, naturally resulting in performance variations proportional to the chosen parameters. Unlike many previous works, this paper presents measurement evaluations in central Auckland, New Zealand, from a \emph{pre-commercial} deployment of a single-user, single-cell 5G-NR system operating in \emph{both} bands. The net throughput, coverage reliability, and channel rank are analyzed across the two bands with baseband and analog beamforming. Our results show that the C-band coverage is considerably better than mmWave, with a consistently higher channel rank. Furthermore, the \emph{spatial stationarity region} (SSR) for the azimuth angles-of-departure (AODs) is characterized, and a model derived from the measured beam identities is presented. The SSR of azimuth AODs is seen to closely follow a gamma distribution. 
\end{abstract}

\vspace{-4pt}
\section{Introduction}
\label{introduction}
\vspace{-2pt}
Deployment of fifth-generation (5G) new radio (NR) systems is taking place from 2019 onward in North America, Europe, Oceania, and parts of Asia. As shown in  Table~\ref{tab_Band_number_for_5G_bands}, there are \emph{five} popular frequency bands for the initial rollout of 5G-NR systems, categorized \emph{below} and \emph{above} 6 GHz carrier frequency. \cite{ITU1}. Well established by now, the dominating 5G use cases can be broadly classified into \emph{three} categories: (1) Enhanced mobile broadband (eMBB), (2) enhanced machine-type communications (eMTC), and (3) ultra-reliability low-latency communication (URLLC) \cite{SHAFI1,ITU3}. Initial deployments of 5G systems will be for the eMBB use case, motivated by the exponentially increasing data rate demands, as well a dramatic increase in the number of user equipments (UEs) per-square kilometer \cite{DAHMAN1,ROH1}. The \emph{two} dominant deployment modes of eMBB systems are \emph{non-stand alone (NSA)}, and \emph{stand alone (SA)}, respectively. Technical specifications for NSA and SA architectures are now frozen by third generation partnership project (3GPP) Release 15 (see e.g., \cite{3GPP1} for a summary). In the NSA system, a 5G-NR cell acts \emph{secondary} to a master long term evolution (LTE-A) cell, and derives its control channel via the LTE-A cell anchor. Here, \emph{both} the 5G and LTE-A systems are connected to the enhanced packet core. In stark contrast, for SA deployments, 5G cells/base stations (BSs) are expected to be directly connected to the next generation core network. Here it is possible that a C-band cell may act as a \emph{master} to a NR millimeter-wave (mmWave) band cell. 

The target of 5G-NR systems is to enhance the system throughput and coverage reliability by an order-of-magnitude relative to contemporaneous LTE-A systems, operating in the C-band. It is well established in theory, and proven by measurements, that the C-band has a coverage advantage relative to the mmWave bands due to the larger \emph{effective area} of antenna arrays and higher efficiency of diffraction \cite{SHAFI1,HALVARSSON1,GENTILE1,MEDBO1}. Furthermore, with the use of massive multiple-input multiple-output (MIMO), a higher \emph{beamforming gain} can be leveraged which enhances the coverage, while resulting in greater throughput capabilities \cite{SHAFI1,SHAFI2,MOLISCH1}. To the best of our knowledge, relatively few \emph{pre-commercial} trials have been reported in the open literature. For instance, authors of \cite{GUAN1} present the throughput performance of a 64 antenna massive MIMO system serving 8 UE co-located UE antennas within 4.5-4.8 GHz. Furthermore, the downlink throughput gains of a 2.6 GHz 128 antenna BS relative to a 8 antenna LTE-A system is presented in \cite{LIU1}, where the beamforming gain of massive MIMO is examined. Coverage predictions of 5G-NR systems at 3.5 GHz are presented in \cite{HALVARSSON1} and \cite{WIRTH1}, while coverage, performance, and beam management concepts are investigated at 28 GHz with pre-commercial setups in \cite{HALVARSSON2,SIMONSSON1,OBARA1}. \emph{Despite this, the majority of the above studies do not identify the fundamental physical layer attributes that primarily differentiate the resulting system performance.}

In order to confirm the coverage and throughput related aspects, a pre-commercial trial of 5G equipment in the C-band (3600-3700 MHz) and in the N257 mmWave band was conducted in downtown Auckland, New Zealand.\footnote{Strictly speaking, we note that band N257 does \emph{not} include mmWave frequencies, i.e., from 30-300 GHz. However, since it contains frequencies which are \emph{approaching} mmWaves (26.5-29.5 GHz), it is loosely referred to as a mmWave band.} This paper presents the measurement results from these trials. The conducted measurements were with a single UE in a single-cell scenario, since our primary focus was to assess and predict the performance of C-band and mmWave systems within the \emph{same} deployment area. \emph{This is largely missing in the related literature (see the above references), where the majority of the studies focus on performance characterization in a single band.} Due to the future inter-workings of 5G-NR systems across the two bands, it is necessary to clearly identify and highlight the physical mechanisms that lead to coverage and throughput differences at both bands. During the course of the deployment, we observe some important differences between the propagation properties, as well as system related aspects. In this paper, we present these differences, and compare with what has been written in the official guidelines for 5G-NR systems \cite{ITU2}. More specifically, our observations from the field deployment consist of the following facts: 
\begin{itemize}
    \item The \emph{sharp shadows} (lack of diffraction) in the mmWave band substantially reduce the radio coverage, in particular behind street corners. The small coverage area is a result of high pathloss and relatively small (in an electrical context) antenna array size. This is in contrast to the C-band, where larger coverage areas are observed.
    \item The instantaneous rank of the propagation channel in the C-band is higher than the mmWave band. \emph{This primarily leads to higher throughput for the C-band in comparison to mmWave for the same radio bandwidth, at the same SNR, within the same area}. 
    \item The \emph{spatial stationarity region (SSR)} (defined later) of the azimuth  angle-of-departures (AODs), determined by monitoring the beam identities (IDs) over the designated measurement area, is gamma distributed when comparing the measured data to a statistical fit. This is contrast to existing models presented in \cite{ITU2}, \cite{KURRAS1}, and \cite{JU1}, respectively.
\end{itemize}
\begin{table}[!t]
\vspace{7pt}
\begin{centering}
\scalebox{0.7}{
\begin{tabular}{ccc}
\toprule 
 & \textbf{3GPP Band \#}  & \textbf{Range}\tabularnewline
\midrule
\midrule 
\multirow{3}{*}{\textbf{Below 6 GHz}} & N77  &  3300\textendash 4200 MHz\tabularnewline
\cmidrule{2-3} 
 & N78  & 3300\textendash 3800 MHz\tabularnewline
\cmidrule{2-3} 
 & N79  & 4400\textendash 5000 MHz\tabularnewline
\midrule 
\multirow{2}{*}{\textbf{Above 6 GHz}} & N257  & 26.5-29.5 GHz \tabularnewline
\cmidrule{2-3} 
 & N258 &  24.5-27.5 GHz\tabularnewline
\bottomrule
\end{tabular}}
\par\end{centering}
\label{tab_Band_number_for_5G_bands}
\vspace{8pt}
\caption{Band numbers for 5G-NR Systems}
\vspace{-25pt}
\end{table}
The rest of the paper is organized as follows: Section~\ref{SystemandMeasurementDescriptions} presents the system and measurement setup descriptions for the deployed systems at both bands. Furthermore, Sec.~\ref{MeasurementResults} presents the measurement results with their relationship to the underlaying physical layer mechanisms. The aforementioned model of SSR is also described and characterized in this section. Finally, Sec.~\ref{Conclusions} concludes the paper.

\vspace{-2pt}
\section{System and Measurement Descriptions}
\label{SystemandMeasurementDescriptions}
Since commercial deployment of 5G-NR aims for a seamless fusion of communications in the C-band and mmWave band, it is vital to study the obtainable coverage and throughput at both bands when transmitting from the same location. Naturally, the transmitter and receiver parameters of both systems are different, due to various implementation aspects. Despite all of the differences between the transmit/receive parameters across the two bands, it is still worthwhile to \emph{cautiously} compare the resulting system performance. The measurements were conducted near the Viaduct harbor are of Auckland, the largest city in New Zealand, using a pre-commercial 5G-NR system. We recognize that the results presented in the paper cannot be generalized to other scenarios, since it is non-trivial to separate out system-related aspects from aspects of the physics governing the system behavior. Nevertheless, our results do provide useful insights into 5G-NR performance within the measured area. 

The system parameters are summarized in Table~\ref{MeasurementParameters}. 
\begin{table}[!t]
\vspace{8pt}
\begin{centering}
\scalebox{0.71}{
\begin{tabular}{ccc}
\toprule 
\textbf{System Parameter} & \textbf{C\textendash Band} & \textbf{mmWave Band}
\tabularnewline
\midrule
\midrule 
\textbf{Frequency Band} & 3.60 --  3.70 GHz &  26.65 -- 27.45 GHz\tabularnewline
\midrule 
\textbf{Center Frequency}  & 3.65 GHz & 27.05 GHz\tabularnewline
\midrule 
\textbf{Bandwidth}  & 100 MHz  & 100/400/800 MHz\tabularnewline
\midrule 
\textbf{EIRP}  & 73 dBm  & 62 dBm\tabularnewline
\midrule 
\textbf{Horizontal 3 dB Beamwidth}   & 13$^{\circ}$  & 22$^{\circ}$ \tabularnewline
\midrule 
\textbf{Vertical 3 dB Beamwidth}   & 6$^{\circ}$  & 4$^{\circ}$ \tabularnewline
\midrule 
\textbf{DL/UL Signalling}  & CP-OFDM  & CP-OFDM\tabularnewline
\midrule 
\textbf{Multiplexing Mode}  & TDD  & TDD\tabularnewline
\midrule 
\textbf{Subcarrier Spacing} & 60 KHz  & 120 KHz\tabularnewline
\midrule 
\textbf{Subcarriers \# Per Resource Block}  & 12 & 12\tabularnewline
\midrule 
\textbf{Maximum \# Resource Blocks}  & 136 & 68 per 100 MHz\tabularnewline\bottomrule
\end{tabular}}
\par\end{centering}
\label{MeasurementParameters}
\vspace{12pt}
\caption{C-Band and mmWave Band Measurement System Parameters}
\vspace{-22pt}
\end{table}
It can be readily observed that within the mmWave band, \emph{multiple} system bandwidths are tested ranging from 100 MHz to 800 MHz. Furthermore, the effective isotropic radiated power (EIRP) is 11 dB higher in the C-band relative to mmWave, while there is a 9$^\circ$ difference in the horizontal 3 dB beamwidth of the antenna arrays used for both bands. The C--band transmitter (BS) uses 192 antenna elements, which form a uniform planar array (UPA) configuration of 8 horizontal and 12 vertical cross-polarized elements. Note that, \emph{zero-forcing (ZF) baseband processing} is implemented for beam generation in the C--band system with 64 radio-frequency (RF) up-conversion chains.\footnote{We note that the inter-UE interference cancellation ability of ZF beamforming is \emph{not} utilized throughout the measurements, since only a \emph{single} UE is present. However, the \emph{inter-stream} interference may be nulled with the ZF processing when multiple streams of data transmitted.} The RF chains are interfaced with 64 power amplifiers (PA) going up to the antenna ports, where each PA is designed to drive 3 antenna elements. In contrast to the C-band, the mmWave transmitter (BS) is equipped with 320 elements in a UPA, which in turn consists of 4 sub-arrays each having 4 horizontal and 20 vertically cross-polarized elements. Unlike the C-band, analog beamforming is employed within the sub-arrays that are used to transmit the four beams. The baseand waveform first passes through a 4 stream multiplexer before being up-converted to the center frequency. In total, there are 16 PAs distributed across the two polarization states in the array. In the vertical polarization, one PA is configured to drive 20 antenna elements and in the horizontal polarization, one PA is configured to drive one element. For both systems, multiple \emph{antenna elements} of a given polarization type are grouped and combined to form a single \emph{logical element}. Due to both system architectures being vendor specific, we refrain from presenting detailed schematic diagrams of the antenna layouts.  
\begin{figure}[!t]
    \centering
    \includegraphics[width=7.5cm]{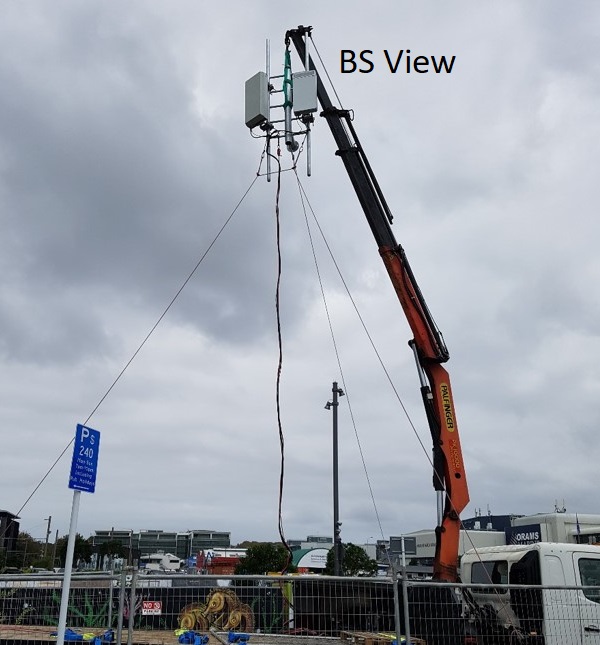}
    \caption{5G-NR BS site deployed at Wynyard central in Auckland, New Zealand. The BS was located at a height of 12 m from the ground level.}
    \label{BSView}
    \vspace{3pt}
\end{figure}
\begin{figure}[!t]
    \centering
    \includegraphics[width=7.5cm]{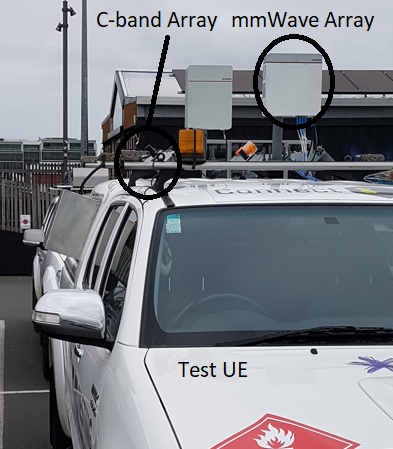}
    \caption{Test UE vehicle deployed in the Auckland measurement area.}
    \label{UEView}
    \vspace{-10pt}
\end{figure}

For the mmWave band, the test UE antenna architecture and pattern consists of four beams which result in an \emph{effective} pattern that is close to omnidirectional in the azimuth plane. Nevertheless, the effective pattern is not completely onmidirectional, since there are nulls \emph{between} adjacent beams that may influence the received signal strength within those angular directions. The C-band test UE antenna pattern is purely omnidirectional in the azimuth plane. Both the C-band and mmWave UEs have a peak broadside gain of 3.2 dBi. Figures~\ref{BSView} and \ref{UEView} illustrate the deployed pre-commercial 5G-NR BS and test UE to conduct the field measurements. As can be observed from Fig.~\ref{BSView}, the BS was placed 12 m above the ground level, while the UE was located on top of the test vehicle in Fig.~\ref{UEView} at approximately 1.5 m above ground level.

\vspace{-4pt}
\section{Measurement Results and Discussion}
\label{MeasurementResults}
\vspace{-1pt}
In this section, the results of the conducted measurements are presented and analyzed to understand the fundamental differences between C-band and mmWave band system performance. Furthermore, we relate aspects of system performance to the underlying differences in physical propagation processes at the two frequency bands. For both systems, the BS global positioning system (GPS) locations are -36.84142778 (latitude) and 174.75526667 (longitude), respectively. On the UE side, a drive test was performed where a test UE was driven in Auckland city around a trajectory (discussed later in more detail). The received data samples are recorded with a sampling distance of 1-3 m. \emph{Multiple} measurement samples were recorded at the same GPS location of the UE due to the drive test vehicle sometimes being \emph{static}. To this end, clustered data samples are \emph{temporally averaged} across the times the test UE vehicle remained static. 

\begin{figure}[!t]
\vspace{1pt}
\includegraphics[width=8.8cm]{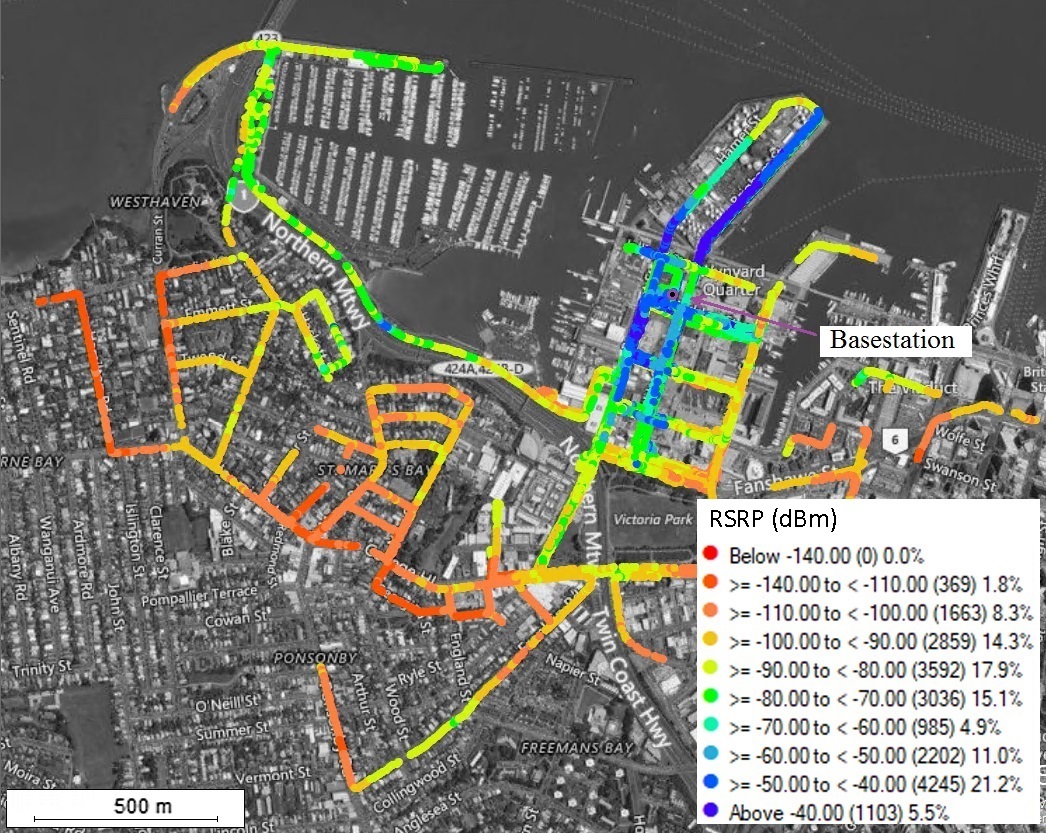}
\caption{C-band geoplot of RSRP measurements in Auckland city.}
\label{GeoplotCband}
\vspace{-2pt}
\end{figure}

\begin{figure}[!t]
\centering
\includegraphics[width=8.6cm]{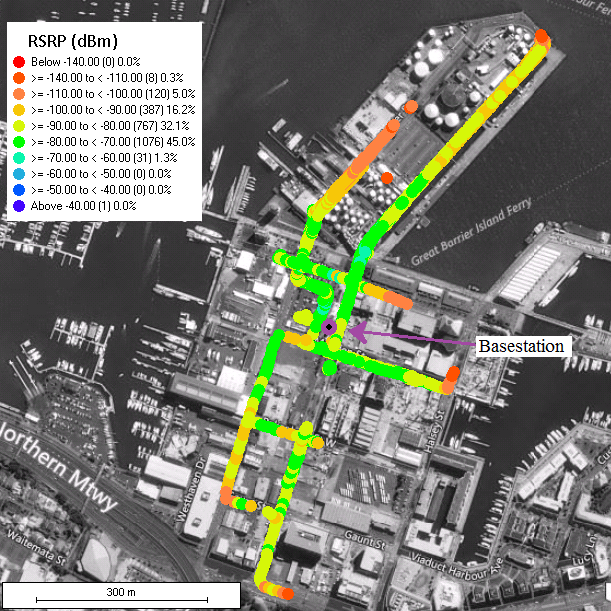}
\caption{MmWave geoplot of RSRP levels around Auckland's Viaduct harbor.}
\label{GeoplotMmwave}
\vspace{-15pt}
\end{figure}

The reference signal received power (RSRP) geo-plots at C-band and mmWave are depicted in Figs.~\ref{GeoplotCband} and \ref{GeoplotMmwave}, respectively. It is noteworthy that the aforementioned results, as well as the subsequent discussions are based on the use of \emph{same} bandwidth of 100 MHz for both systems. As can be observed, the C--band measurements are conducted over a \emph{larger} geographical area relative to the mmWave measurements. This is to be expected due to lower propagation losses at C-band, and higher EIRP. The maximum link distance for C--band and mmWave band measurements are 1730 m and 640 m, respectively. When comparing the RSRP levels across the two figures, we can see that in the mmWave case, the RSRP \emph{decreases severely} when the UE moves around street corners. Such an effect occurs multiple times, and is to be expected for several reasons: Firstly, it is natural to point out that higher layer decisions of both systems will contribute to this.\footnote{Since pre-commercial systems were used, we avoid further discussions on these decisions, due to vendor specific design and implementation.} Secondly, the physical propagation processes (contributing to the channel characteristics) which are more pronounced in the C-band are no longer dominant at mmWave frequencies. In particular, the first Fresnel zone, which includes in excess of 98\% of the propagating energy becomes very narrow at mmWave due to the wavelength being close to 10 mm (decreasing as square root of wavelength). Due to this reason,  \emph{diffraction} becomes highly inefficient and multipath components have higher blockage probability, since common obstacles such as building walls, cars and human beings introduce sharp shadows. In the C-band (as well as in canonical systems of today), diffraction plays a vital role in achieving the desired coverage. Thirdly, the magnitude of \emph{reflections}, which are caused by building walls as the test UE moves around street corners depends greatly on the operation wavelength, and thus on the carrier frequency. Since reflection and transmission happen \emph{simultaneously}, power transmitted through the walls \emph{decreases} almost uniformly with carrier frequency due to the presence of the skin effect in lossy media, as well as the \emph{electrical thickness} of walls. In addition, one can observe that in the mmWave case, the received signal to the UE is entirely blocked towards the top left-right hand corner of the geo-plot, where the system is unable to record RSRP levels. \emph{The reasons for this are discussed in the next paragraph.}

\begin{figure*}[!t]
\centering
\includegraphics[width=17.5cm]{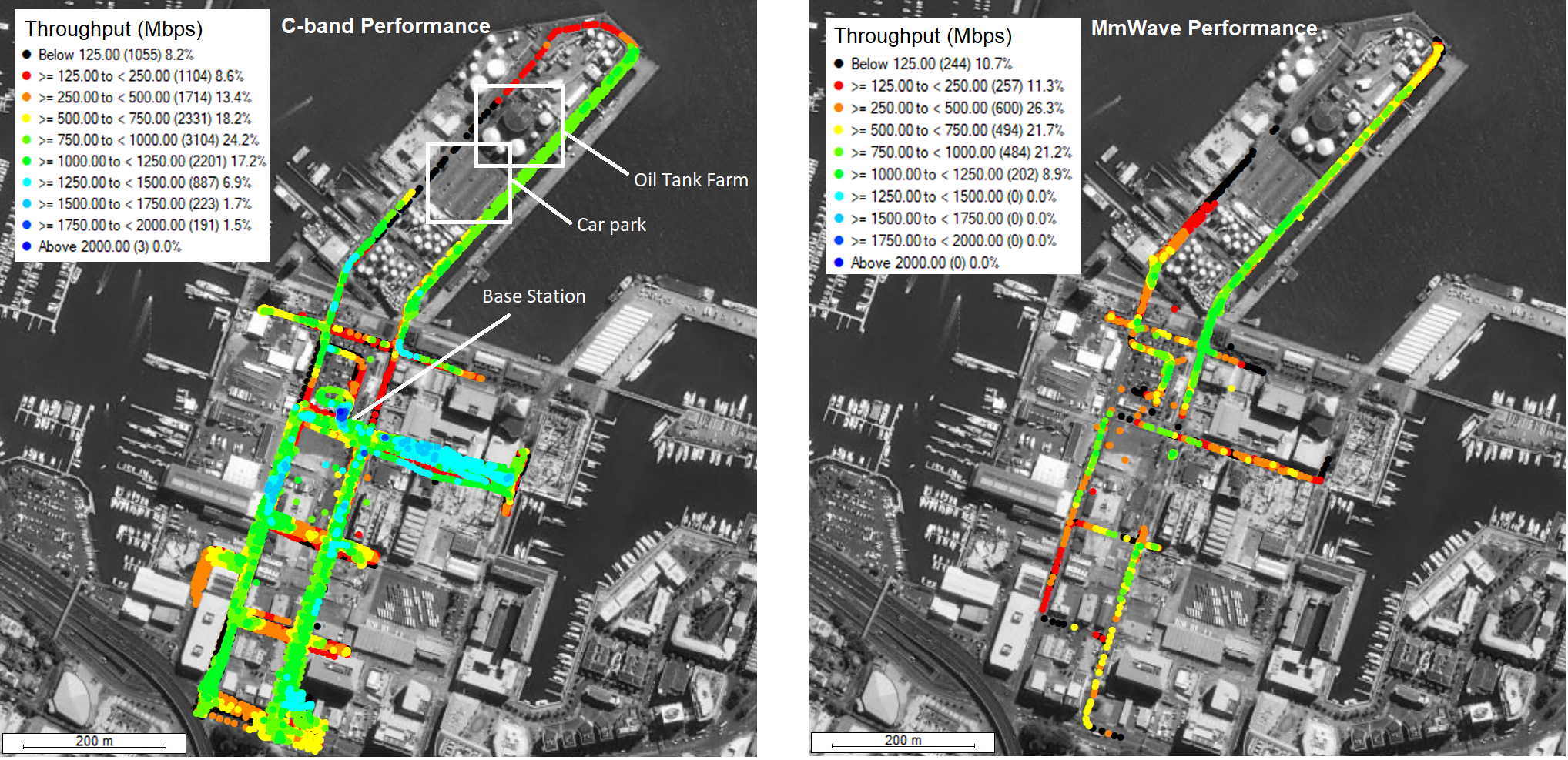}
\caption{\emph{Left-hand side}: C-band throughput performance at 100 MHz. \emph{Right-hand side}: MmWave throughput performance in the \emph{same area} at 100 MHz.}
\label{SameAreaTPerf}
\end{figure*}

Since the physical area of the conducted trials is larger for the C-band relative to mmWave, in what follows, for the sake of fair comparison, we present measurement evaluations over the \emph{same} geographical area for both systems at 100 MHz. To this end, Fig.~\ref{SameAreaTPerf} (see top of the following page) demonstrates the downlink throughput performance of the two systems. While similar overall conclusions can be retained from the RSRP results earlier, we are able to clearly observe the throughput differences between C-band and mmWave. Due to the high diffraction losses at mmWave, several locations can be identified where the throughput reaches a level \emph{below} the minimum requirement where the UE is unable to receive and record data. Two prominent instances of such events can be identified as the test UE moves behind the \emph{car park} and \emph{oil tank farm} regions, identified with white boxes in the left-hand side sub-figure. The large metallic obstructions caused by oil tankers and car bodies block the 28 GHz signal completely. Interestingly, the blockage effects are still present at C-band, where the throughput is seen to go below 125 Mbps (black region). Nevertheless, it is clear that the C-band provides better throughput across the measurement area relative to mmWave. 
\begin{figure}[!t]
\begin{centering}
\vspace{-3pt}
\includegraphics[width=8.65cm]{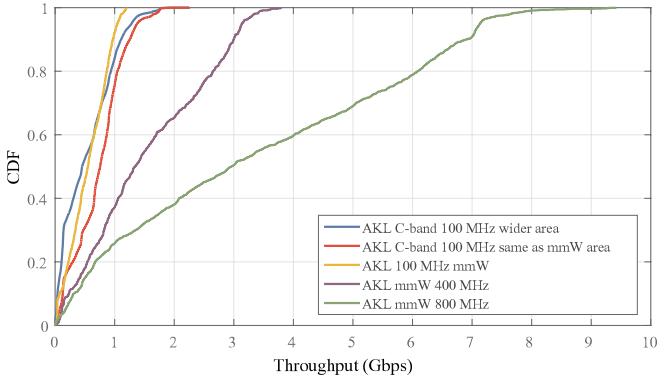}
\par\end{centering}
\vspace{-5pt}
\caption{Measured C-band and mmWave throughput CDFs for system bandwidths of 100, 400 and 800 MHz.}
\label{Fig_CDF_TP_AKL_mmW_Cband_TP}
\end{figure}

\vspace{-12pt}
We now turn our attention to the variation in performance of both systems with different system bandwidths. Figure~\ref{Fig_CDF_TP_AKL_mmW_Cband_TP} presents the measured cumulative distribution functions (CDFs) of the downlink throughput for C-band 100 MHz, and mmWave band 100, 400, and 800 MHz bandwidths. The throughput CDFs for C--band measurements over the same area as the mmWave measurements are also given in the figure for fair comparison. The ensemble over which the CDFs is taken are the UE positions, which map onto the received downlink signal-to-noise ratios (SNRs). Several important trends can be observed: Firstly, it can be noted that the C-band throughput is \emph{higher} than the corresponding mmWave throughput for the \emph{same} bandwidth, across the same area (comparison of red and yellow curves). More specifically, an increased throughput of approximately 0.25 Gbps at CDF = 0.5 and 0.3 Gbps at CDF = 0.9 is observed. These values are representatives of the \emph{median} and \emph{peak} throughput differences between the two systems. Secondly, around 20\% of the total captured measurements at the C-band are above 1 Gbps, whereas only around 5\% of the measurements above 1 Gbps are recorded in the mmWave band. Thirdly, the median throughput seems to \emph{double} as the bandwidth grows from 100 to 400 and 800 MHz. On a similar line, increasing the bandwidth from 100 to 400 MHz yields a 2 Gbps increase in the peak throughput (see CDF = 0.9), while a 4 Gbps increase is noticed as the bandwidth increases yet again from 400 to 800 MHz. Rather interestingly, the upper tails of the CDFs broaden out relative to the narrow and confined lower tails, indicating a large degree of right-hand side skewness in the underlaying probability densities. Returning back to the throughput superiority of C-band relative to mmWave, many design specific factors contribute to this, such as the link level scheduler decision making. In addition, the \emph{channel rank} is higher for the C-band relative to mmWave, as measured by the rank indicator of the system. The \emph{maximum} number of data streams transmitted to a single UE is 8 for the C-band, and 4 for mmWave.
\begin{table}[!t]
\begin{centering}
\vspace{5pt}
\scalebox{0.78}{
\begin{tabular}{ccccccccc}
\toprule 
\textbf{Rank} & 1 & 2 & 3 & 4 & 5 & 6 & 7 & 8
\tabularnewline
\midrule
\midrule 
\textbf{Occ. Prob.}  & 0.32 & 0.2 & 0.17 & 0.23 & 0.03 & 0.02 & 0.01 & 0.02 \tabularnewline\bottomrule
\end{tabular}}
\par\end{centering}
\label{CbandRank}
\vspace{10pt}
\caption{C-band rank occurrence probabilities for the route in Fig. 3.}
\vspace{-17pt}
\end{table}

\begin{figure}[!t]
\vspace{-9pt}
\hspace{-15pt}
\includegraphics[width=9.8cm]{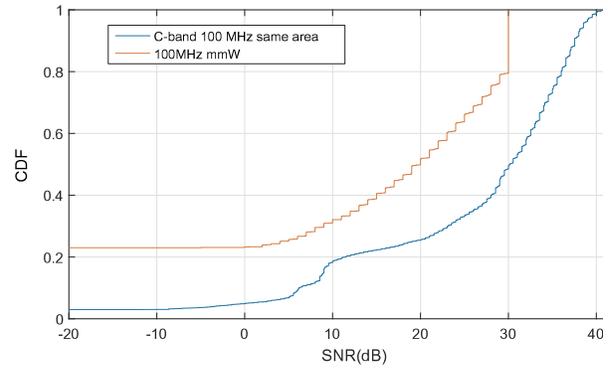}
\vspace{-12pt}
\caption{Measured C-band and mmWave SNR CDFs for 100 MHz bandwidth over the same measurement area.}
\label{SNRCDFs}
\vspace{-9pt}
\end{figure}

On average, our observations show that the mmWave channel does not support a rank of more than 2. Consequently, the performance of 4 streams is inferior to that of 2 streams due to the inter-stream interference that arises from the inability of the channel to support 4 independent streams (no receive processing is used). Also, the mmWave system architecture does not have the ability to mitigate inter-stream interference, since it is implemented using analog beamforming. The occurrence probabilities of the rank for the C-band system are given in Tab.~\ref{CbandRank}. As can be readily observed, a maximum rank of 8 is achievable. This is since a part of the UE route goes through the Auckland central business district which contains medium-rise buildings of 10-12 floors, creating clusters of reflection points. In addition, approximately 50\% of the time, the rank is grater than 2, contributing to the performance superiority in comparison to the mmWave system. The CDFs of underlaying SNRs\footnote {Note that the SNR denoted here is the \emph{reference element} SNR reported by the UE and measured by the system on its allocated sub-carrier.} for C-band and mmWave band over the same measurement area is shown in Fig.~\ref{SNRCDFs}. Note that the SNRs are impacted not only by the EIRP and channel attenuation, but also beamforming type, antenna array sizes, and power control algorithms, respectively. 
Due to the saturation of mmWave measurement equipment, the SNR values \emph{above} 30 dB are recorded and capped as 30. This is the reason for the vertical line in the red curve from CDF = 0.8. At the lower tail of the CDFs, very low SNR levels are seen. This is due to the UE positions where the RSRP levels were seen to drop below the minimum threshold, where the transmitted data streams/packets could not be successfully decoded. Examples of such locations are around the tank farm and car park areas indicated in the left-hand sub-figure of Fig.~\ref{SameAreaTPerf}. While this effect is more prominent in the mmWave relative to C-band, lower SNR values are also recorded in the C-band around these regions. To have a fair comparison of the SNRs between the two bands, the percentage of \emph{missing} data packets is calculated by analyzing the missing GPS timestamps. Since approximately 23\% of the data was lost in the mmWave band and 3\% in the C-band, the CDFs for mmWave and C-band effectively begin at 0.23 and 0.03. Nevertheless, for the \emph{same} CDF value, we can observe a 10 dB SNR gain for the C-band relative to mmWave.\footnote{These conclusions naturally come with a \emph{cautionary tale} due to the involvement of higher layer processing and different power control algorithms for both systems.}

In the following results, we present the measurement-based evaluation and modelling of SSR. While the exact definition of SSR is described later in the text, we note that SSR is closely related to \emph{spatial consistency} (SC) \cite{RAPPAPORT1}, which is a novel feature of 3GPP 38.901 channel model \cite{3GPP2}. The main idea behind SC is that propagation channel parameters such as delays, angle-of-arrivals (AOAs) and AODs should evolve \emph{continuously} as the UE moves along a trajectory. This is in stark contrast to the canonical \emph{drop-based} models in which channel parameters between multiple UE ``drops" can change abruptly, without any relation to the UE location. Here, our objective is to derive a simple model for the SSR of azimuth AODs.\footnote{We recall that since the UE has an approximately omnidirectional radiation pattern, it is difficult to characterize AOA with a reasonable accuracy.} \emph{We define the SSR region as the distance over which the UE is served by the same beam from the BS array.} Taking the example of the mmWave system, there are 4 transmit beams with approximately 20$^{\circ}$ 3 dB beamwidth in azimuth, having \emph{central} angles of 
0$^{\circ}$, $-$24$^{\circ}$, $+$24$^{\circ}$, and $-/+$48$^{\circ}$, respectively. Naturally, the transmitted data streams are sent to the UE via one of the four possible beams. As the UE moves from one position to the next, the BS selects the ``optimal" beam to transmit on, computed by a vendor specific algorithm. For each location of the UE, the system measures and records a beam ID, which is associated to a beam number between 0 and 3. The beam ID serves as a proxy for estimating AOD SC, and hence SSR. 
\begin{figure}[!t]
\vspace{3pt}
\hspace{-5pt}
\includegraphics[width=8.3cm]{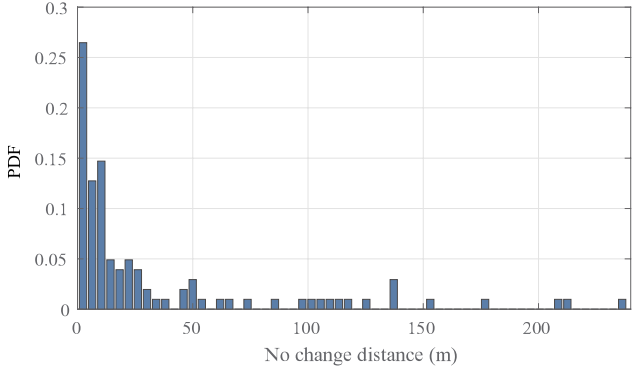}
\vspace{-4pt}
\caption{PDF of SSR distance for the mmWave system at 100 MHz bandwidth.}
\label{PDFBeamNoChange}
\vspace{-10pt}
\end{figure}
\begin{figure}[!t]
\includegraphics[width=8.1cm]{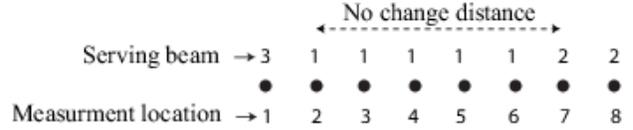}
\vspace{-2pt}
\caption{An illustration of no beam change distance calculation.}
\label{SSDistComputation}
\vspace{5pt}
\end{figure}
In Fig.~\ref{PDFBeamNoChange} we show the probability density function (PDF) of the SSR distance (labeled as ``No change distance" in the fig.) for the 100 MHz mmWave system. We compute the SSR distance as the distance of separation between consecutive measurement points over which there is \emph{no beam change}. An illustration of how the no beam change distance is calculated is shown in Fig.~\ref{SSDistComputation}. According to the literature, SC measurements are still in their infancy, and ideally further measurements are necessary to model SC across different environments. This is since the SSR distance will be different for different environments. According to \cite{KURRAS1}, a variety of statistical distributions can be used to stochastically model the SSR. We employ the widely used gamma distribution, which belongs to the family of two-parameter distributions to model the measured SSR density. Our choice was governed by the distribution which resulted in the highest Kolmogorov-Smirnov test statistic across a wide range of two parameter distributions, such as log normal, gamma, and beta distributions, respectively. The measurement data was shown to lie within a 95\% confidence interval. The PDF of the gamma distribution is given by \cite{SUN1,TATARIA1} 
\vspace{3pt}
\begin{equation}
    \label{gammapdf}
    f(x)=\frac{\beta^\alpha}{\Gamma\left(\alpha\right)}x^{\alpha-1}
    \exp^{-\beta{}x};\hspace{5pt}x>0, 
    \vspace{3pt}
\end{equation}
where $\Gamma(\alpha)$ is the scalar gamma function defined for all positive values of $\alpha$. Note that $\alpha$ and $\beta$ are the \emph{shape} and \emph{rate} parameters of the distribution, which in our case are characterized by $\alpha=0.62$ and $\beta=55.6$ for the tightest point-wise fit. 
\begin{figure}[!t]
\includegraphics[width=8.3cm]{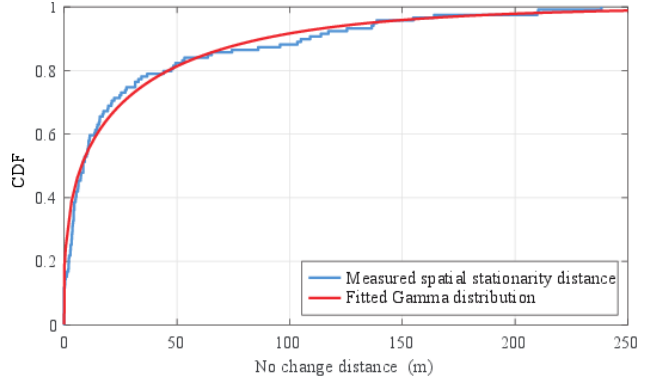}
\vspace{-4pt}
\caption{SSR CDFs for the measured 100 MHz mmWave system and its corresponding gamma distribution model.}
\vspace{-10pt}
\label{CDFSSWithGamDist}
\end{figure}
Figure~\ref{CDFSSWithGamDist} depicts the CDFs of the SSR and the fitted gamma distribution, respectively. We observe that the \emph{median} distance over which the transmit beam remains the \emph{same} is approximately 10 m. Likewise, for 90\% of the distance over which there is \emph{no change} is less than 125 m. Similar trends were observed for mmWave measurements with 400 and 800 MHz bandwidth. At a given distance $d$, the mean azimuth AOD, denoted by $\phi(d)$, takes on one of four values in $\phi_0$, $\phi_1$, $\phi_2$ and $\phi_3$ respectively. When the UE location changes, SC will impose a restriction on how $\phi(d)$ will evolve from its current value to other values, $\phi\left(d+d_0\right)$, at distance $d+d_0$. 

In the case when there is a beam change, Tab.~IV shows that the \emph{median} beam change is to the \emph{first} adjacent beam. However, there are also changes also to the second and third adjacent beams, respectively. One can observe that there is a higher likelihood of a beam change to the third adjacent beam than the second adjacent beam. We conjecture that this may be attributed to the scattering cluster locations in the measurement area, however further measurements are required to accurately validate this conjecture. 
\begin{table}[!t]
\label{BeamChangeProb12}
\vspace{10pt}
\begin{centering}
\scalebox{0.78}{
\begin{tabular}{cccc}
\toprule 
\textbf{Beam Change} & First Adjacent & Second Adjacent & Third Adjacent\tabularnewline
\midrule
\midrule 
\textbf{Occ. Prob.}  & 0.63 & 0.06 & 0.31 \tabularnewline\bottomrule
\end{tabular}}
\par\end{centering}
\vspace{9pt}
\caption{Beam change probabilities for the 100 MHz mmWave system.}
\vspace{-22pt}
\end{table}
Keeping in mind the aforementioned discussion, the probability that the beam change is to the \emph{first} adjacent beam can be expressed as
\begin{align}
    \nonumber
     &\mathbb{P}\left[\hspace{1pt}\phi\hspace{2pt}(d+d_0)=\phi_1\right]=
     \frac{1}{\mathbb{P}\left[\phi\hspace{2pt}
     (d+d_0)\neq\phi_0\hspace{1pt}\right]}\\[8pt]
    \label{firstbeamchange}
     &\times\mathbb{P}\left[\hspace{1pt}\phi\hspace{2pt}(d+d_0)=\phi_1 \text{ and } \phi\hspace{2pt}(d+d_0)\neq\phi_0\hspace{1pt}\right].\\[-14pt]
     \nonumber
\end{align}
Following very similar calculations, the probabilities of beam change to the second and third adjacent beams can also be obtained. In contrast to our approach, which characterizes the azimuth AOD SSR as a gamma distributed random variable, it is modelled as a \emph{deterministic} quantity in \cite{ITU2,KURRAS1,WANG1}. Further measurements are needed to make remarks about the generality of this conclusion under different environments.

\vspace{-3pt}
\section{Conclusions}
\label{Conclusions}
This paper has provided a comparison between C--band and mmWave  system performance based on pre-commerical 5G-NR field trials. The comparisons of RSRP and throughput over the same area has indicated that the mmWave band has lack of coverage and is more sensitive to blockages, especially around the street corners. Although the system parameters for both bands were not the same (due to differences in the underlaying design decisions), the lack of coverage can be most likely attributed at the physical layer to the inefficiency of diffraction at mmWave frequencies. The channel rank and SNR in the C-band was observed to be higher than the corresponding mmWave values. This has also translated to the increased throughput for 100 MHz C--band measurements for the same area in comparison to the mmWave band measurements. The SSR of azimuth AODs of the mmWave channel at 100 MHz is derived from the beam IDs. This indicated that the median distance over which the transmit beam remains the same is about 10 m, implying that the mean azimuth AODs are fully correlated across 10 m. Further to this, it was demonstrated that the azimuth AOD SSR can be modelled as a gamma distributed random variable, where its shape and rate parameters were characterized. A point wise fit was shown to tightly agree with measurement evaluations.

\vspace{-6pt}
\section*{Acknowledgements}
\vspace{-3pt}
The authors gratefully acknowledge Dr. Sujith Ramachandran for his help with data processing 
of the measurements.


\end{document}